# *Buried Moiré supercells through SrTiO$_3$ nanolayer relaxation*

*Max Burian†,\*, Bill Francesco Pedrini†, Nazaret Ortiz Hernandez†, Hiroki Ueda†, C. A. F. Vaz†, Marco Caputo†, Milan Radovic† and Urs Staub†,\**

† Swiss Light Source, Paul Scherrer Institute, 5232 Villigen PSI, Switzerland

\* To whom correspondence should be addressed. Email: urs.staub@psi.ch and max.burian@psi.ch

Here, we identify a highly ordered Moiré lattice at the deeply buried SrTiO$_3$ (STO) - LSAT interface. Using high-resolution X-ray diffraction reciprocal space mapping, we find long-ranged ordered supercells of 106/107 unit cells of unstrained STO/LSAT, caused by complete lattice relaxation through high-temperature annealing. Model calculations confirm the experimental scattering phenomena and show that cross-interfacial bonding is locally different at the Moiré-overlap points. The presence of such super-ordered structures in the family of 2D electron gas systems sets the ideal conditions for Moiré-motif tuned plasmonic responses and ferroelectric super-crystallinity, opening up the possibility for novel interface functionalities in these simple perovskites and impacts findings on long wavelength ordered vortex structured perovskite multilayers systems.

Complex-oxide heterostructures with their unique set of physical and chemical properties are of looming interest for magneto-electronic devices [1–8]. Here, intrinsic material properties, such as (i) rich spin-interactions [9], (ii) strong multiferroric character [10] and (iii) high thermal and operational stability [11], are extended by electronic interactions at the hetero interface [8,12–14]. Indeed, accurate interface engineering has led to a new functional repertoire [15], including observations of quasi-2D electron gas (q-2DEG) [14,16,17], charge writing [18], resistance switching [19], occurrence of electronic [20] and magnetic order [21], giant thermoelectric effect [22] and colossal ionic conductivity [23]. The origins of these collective phenomena lie in site-specific spin and charge interactions [24–26] that are in turn governed by the local atomic arrangement [27]. Indeed, simple misalignment of two equal lattices induces a long-range ordered superstructure termed "Moiré-motif" that manifests itself in new collective properties, such as e.g. plasmonic minibands [28] or Moiré-excitons [29]. Precise knowledge of the atomic structure at the interface, even between seemingly equal lattices, is hence key in understanding and enhancing the collective material-system behavior.

Prominent candidates for such oxide-based heterostructures are perovskite titanates, including $SrTiO_3$ (STO). The strong interest in STO stems from (i) its involvement in the first report of an interfacial q-2DEG [14] that is strongly linked to (ii) its high and tunable dielectric constant [30,31]. Quantum fluctuations of the cell-centered Ti-atom suppress the paraelectric to ferroelectric transition [32], whereas these fluctuations are driven by the STO soft phonon-mode [33]. As a result, the phononic landscape determines the dielectric anisotropy and magnitude of the material. Lattice straining effects that drastically alter the respective phonon modes therefore play a crucial role in heterostructure growth of STO, leading either to (i) a polar structure [34,35] or to (ii) suppression of lattice polarizability and strong reduction in dielectric character [31,36]. Obtaining ideal bulk-material properties in STO films hence requires perfect relaxation of the crystalline lattice by structural decoupling from the substrate. Indeed, this lattice relaxation might also crucial in the occurrence of ferroelectric 3D vortex structures [37,38], room-temperature skyrmion structures [39] or laser-induced super-crystallinity [40] in STO based nanolayer heterostructures – yet the structural origin of these phenomena is not fully understood.

Here, we look at the interface between a nanolayer of STO grown on a (La,Sr)(Al,Ta)-oxide (LSAT) substrate - a material combination with minimal potential strain: both compounds have a cubic lattice [$a_{STO}$ = 3.904 Å [41]; $a_{LSAT}$ = 3.869 Å [42]] and a mismatch of only +0.93%. Despite this small lattice mismatch, straining from lattice matching of the grown STO layer is sufficient to initially suppress the soft phonon-mode [32,43]. As previously shown one can recover the bulk-like phononic landscape by high temperature annealing in ambient atmosphere [32,43], suggesting structural relaxation of the STO nanolayer. We follow this approach and use a ~50 nm thick film of STO on LSAT [001] grown by pulsed laser deposition, followed by 12 h of annealing at 1200°C at ambient conditions [43]. We find that this heat-treatment leads to complete relaxation of the STO nanolayer, resulting in a Moiré pattern at the STO/LSAT interface. Comparison with theoretical calculations shows that this Moiré interference only occurs if the atomic bond-length at the interface varies at the overlap points. Our explanation not only help to understand recent reports of 3D vortex structures [37–40], but also implies a long-range ordered modulation of the plasmonic nature of the interfacial electron gas [44], opening new applicative perspectives for plasmon tunable complex oxide devices.

X-Ray diffraction (XRD) experiments were performed at the surface X-Ray diffraction (SXRD) endstation of the MS beamline [45,46] at the Swiss Light Source. The X-Ray beam (12.65 keV) was focused and cut to a beamsize of 500 x 500 μm (H x V) at the sample, which was mounted on a (2+3)-type surface diffractometer (vertical geometry) [47] with an angular resolution of 0.002°. During scans, the primary intensity of the incident beam is automatically adjusted (by a variable transmission-filter array) to optimize the effective dynamic range of each scan. The UB "orientation-matrix" describing sample alignment in the beamline coordinate system was defined by manual alignment of 3 non-orthogonal and 2 orthogonal reflections, resulting in a UB accuracy of h,k,l ± 0.0005 r.l.u.. Reciprocal space scans (200 images, 1 s each) over each reflection were made along the (001) and (112) directions to check for possible distortion/misalignment effects. Angular to reciprocal space conversion of each detector pixels was done according to Schlepütz, et.al. [47]. For each scan, an array of all 3D reciprocal space coordinates with the corresponding scattering intensities were interpolated onto an orthogonal, equidistant (voxel size h,k,l = 0.0005 r.l.u) 3D matrix, constituting the reciprocal space volume (RSV). The convolution product of (i) experimental accuracy and (ii) the RSV voxel size yields the effective resolution of the RSV, which in this case is h,k,l ± 0.0007 r.l.u.. A detailed description of the Fast Fourier Transform (FFT) model calculations can be found in the Supporting Information – section 2D and 3D Model Calculations.

As structural motifs of the hetero-interface are often projected onto the sample surface [48], we characterize the STO nanolayer morphology by atomic force microscopy (AFM). As seen in Fig. S1a, AFM shows disordered islands with 5-8 nm thickness and 146 nm mean lateral correlation length (see Figure S1b and c for Fourier-analysis of the AFM image). Such island-growth has been linked to segregation effects during annealing and is usually found to seed at lattice-defects, such as dislocations and/or steps [48–51]. The absence of a long-range ordered island arrangement (here evidenced by disorder) hence points towards an isotropic STO nanolayer morphology decoupled from the crystallographic lattice.

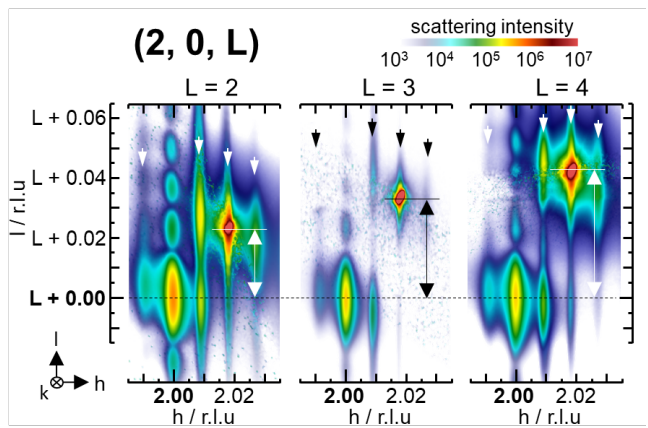

**Figure 1.** Slices through the reciprocal space volumes (RSV) obtained by X-Ray scattering at the (2, 0, L) family peaks. The reciprocal-space distance between substrate and surface-layer peak increases linearly with L-diffraction order, yielding an out-of-plane lattice mismatch of 1.07±0.10 %. Color and transparency are log-scaled with scattering intensity according to the color bar.

To obtain information on the atomic lattice strain state, we collect reciprocal space slices (scattering images) by synchrotron X-Ray diffraction, which we map and assemble into a 3D reciprocal space volume (RSV). The final RSV is transferred into the STO-HKL space [named indices relate to STO and not to the LSAT substrate where capital indices (e.g. H, K or L) denote reference positions in reciprocal space while lower-case indices (e.g. h, k or l) denote scanning variables]. First, we select the low-symmetry (2, 0, L) family with L = 2, 3 and 4, providing information on the *out-of-plane* strain state of the STO film. As shown in Fig. 1, we observe rod-like scattering features extending along the l-axis, typical of an out-of-plane nanolayer. For all Bragg-peaks, oscillations at k=0 along l can be observed, corresponding to layer-thickness dependent fringes (see rocking-curve in Fig. S2a, confirming a STO film-thickness of $54 \pm 2$ nm). More evidently, the RSV presents two significant scattering contributions along the l-direction, from: (i) the layer (at lower l) and (ii) the substrate (at higher l). From the distance between the two scattering centers, we calculate the difference in out-of-plane d-spacing of STO and LSAT, corresponding to a mismatch of $+1.07 \pm 0.10$ %. This value is only slightly above the theoretical prediction for a perfectly relaxed lattice (+0.93%).

Interestingly, the scattering-rods seen as streaks along the out-of-plane l-direction (see Fig. 1) present coherent repetitions along the in-plane k-direction. Note, that the small distance between these rods in reciprocal space must relate to a large origin in real space. Indeed, such scattering features are characteristic of a large-scale and long-range order as found in superstructures [20,52–54], here relating to an in-plane d-spacing of 41.1 nm (corresponding to the reciprocal-space distance of $\Delta k = 0.0096$ between the modulations – see Fig. S2 and S3 for peak-fits). Interestingly, the scattering rods appear to be commensurate with the atomic LSAT and STO Bragg-reflections, as both peaks (of an H = 2 type reflection) are linked by exactly two scattering-rods.

To understand if there indeed exists a relation between the scattering rods and the underlying atomic lattices, we vary the in-plane index within the same (2, K, 4) family, as shown in Fig. 2a. Remarkably, with varying k between -2 to 2, the substrate peak (at l = 4.045) and the STO peak (at l = 4.00) always fall on superstructure scattering-rods. E.g., at the (2, -2, 4) reflection, the substrate peak lies on the 2[nd] scattering rod from STO, whereas increasing k to -1 translates the substrate peak to the neighboring rod. This effect becomes even clearer when looking at orthogonal projections that show the full in-plane relation (see Fig. 2b). At all reflections, the superstructure-rods form an in-plane 2D lattice with square symmetry (see Fig. S2 comparison of cuts along h and k trajectories). Along h, substrate and layer peak are always spaced two scattering-rods apart (note, that H=2 for this reflection), while the substrate peak varies along k on the 2D scattering–rod-grid with increasing diffraction order. This commensurate relation between (i) atomic reflections and (ii) scattering-rods therefore suggests a structural relation between (i) the atomic crystalline lattices and (ii) the formed superstructure.

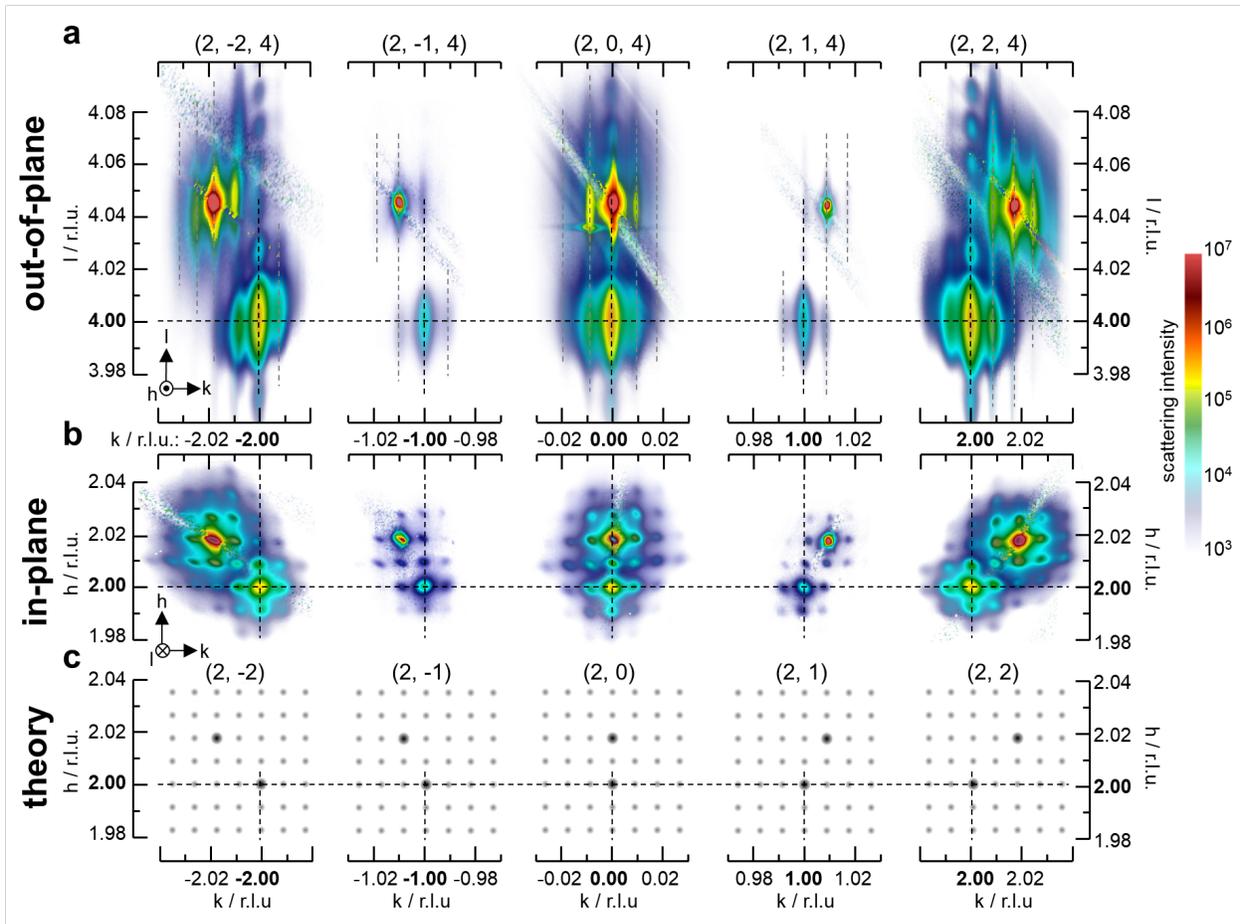

**Figure 2.** **(a)** Out-of-plane and **(b)** in-plane sliced projections of the reciprocal space volume (RSV) obtained by X-Ray scattering on the (2, K, 4) family peaks, showing a commensurate relation between atomic lattice reflections and the superstructure scattering-rods. Black lines mark the center of the Bragg-

reflection in the STO reference frame; grey lines are guides to the eye. Color and transparency are log-scaled with scattering intensity according to the color bar. **(c)** The intensity of 2D Fourier-transform of a theoretical 106/107 Moiré lattice reproduces the same scattering behavior, only if a corrugation induced phase-shift is taken into account.

In regard of the origin of the observed scattering rods, it is important to realize that the superstructure scattering pattern is seen on both the atomic substrate (LSAT) and layer (STO) peak, implying that the underlying structural feature originates from the LSAT/STO interface. [We can rule out surface topology as origin of the observed scattering rods, as we find a coherent domain size of >300 nm (see Fig. S3 for corresponding peak fits), which is significantly larger than the surface features with approximately 150 nm identified by AFM (see Fig. S1).] Further, the ratio between the STO lattice constant (3.904 Å) and the observed superstructure d-spacing (42.5 nm) corresponds to 0.92%, ideally matching the theoretical STO/LSAT lattice-mismatch of 0.93%. The observed scattering behavior hence stems from scattering interference of LSAT and the perfectly in-plane relaxed STO layer, classified as a Moiré-like motif [28,55,56].

In short, such Moiré motifs appear through superposition of two (in this case square) base-lattices with different periodicity $a_{small}$ and $a_{large}$, such that spatial interference gives rise to a new pattern with the same symmetry but a lattice constant of $a_{moire} = (a_{large} * a_{small})/(a_{large} - a_{small})$. [57–59] Here, these two base-lattices relate to STO and LSAT, where it takes 106 unit cells of STO and 107 unit cells of LSAT to compensate the lattice mismatch of 0.93%, yielding a Moiré lattice constant of 40.8 nm. Indeed, and as shown in Fig. 3c, a 2D Fourier transform of such an idealized 106/107 Moiré lattice shows the same scattering motif as observed experimentally: the two strong reflections correspond to the original 106- and 107-base-lattices whereas the weak and smaller-spaced reflections correspond to the Moiré interference terms [see Supporting Information – section Model calculations for a discussion on the origin of such superlattice reflections]. Noteworthy, the here found periodicity of a 106/107-type Moiré-motif is 3-4 times larger compared to reported (few-layer) Moiré-systems [55,56,59–61].

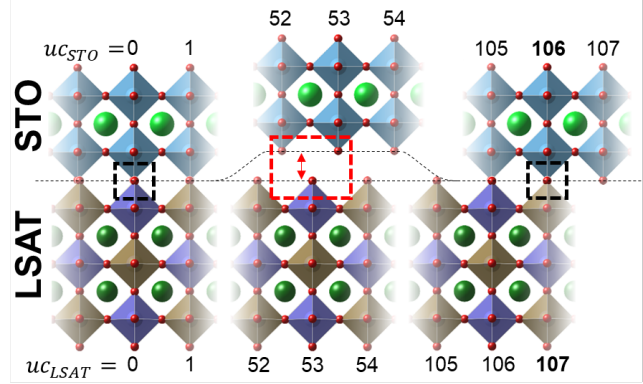

**Figure 3.** Structural illustration of the STO/LSAT interface due to lattice relaxation, shown along the STO (100) direction. Starting from perfect overlap of the corresponding unit cells (uc) [see black dashed area at $uc_{STO}$=0], the local mismatch of the interfacial oxygen atoms increases towards, inducing corrugation of the STO layer at the interface [see red dashed area at $uc_{STO}$=53]. Eventually, the LSAT and STO unit cells overlap again [see black dashed area at $uc_{STO}$=106], effectively forming a 2D 106/107 Moiré pattern.

The observation of this Moiré pattern allows us to draw conclusions about the LSAT/STO interface. First, it is evidence of a perfectly in-plane relaxed STO layer without remnant substrate-induced strain. Second, just few of the STO $TiO_3$-octahedra within a 106 x 106 square fall directly on-top of LSAT $(Ta;Al)O_3$-octahedra (see Fig. 3). Only within the vicinity of this overlap, shared bonding of the STO/LSAT interface oxygen atoms is feasible (see black-dashed area in Fig. 3). Indeed, when calculating the diffraction intensity of such a Moiré pattern, superstructure-reflections only become visible if the scattering amplitude of the overlapping unit cell is phase-shifted compared to the non-overlapping majority [see Fig. S4, S5 and Supporting Information – section 2D Model Calculations for details]. In the sample, this phase shift is a consequence of different out-of-plane bond-length that usually occurs in Moiré lattices, where the related atomic-displacement creates a unique chemical surrounding at this overlap point [61]. Indeed, 3D FFT model calculations show that a repetitive out-of-plane corrugation causes such an out-of-plane phase shift and hence the observe scattering patterns [see Fig. S6 and Supporting Information – section 3D Model calculations for details]. Third, the remaining vast majority of interfacial oxygen atoms (see red-dashed area in Fig. 3) likely do not coordinate in a long-range order, as we do not find any other higher-order scattering terms in the RSVs. In quantitative terms (and assuming a critical local strain of ±3%), less than 0.5% of oxide-octahedra at the interface bond coherently. This strongly suggests (i) an overall disordered interface layer that is (ii) prone to local structural defects at the interface. That this is a property of STO and not the substrate is shown by depositing a thin STO film on $NdGaO_3$, which shows after the exact annealing step the same type of superstructure, but with different lattice spacing due to the different lattice mismatch (see Fig. S7).

The concept of Moiré-motifs also helps to understand more complex structural arrangements, spanning over three dimensions. In a series of works [37–40], X-Ray superstructure reflections similar to the ones shown in this work have been linked to the formation of 3D vortex structures in the heterostructured system of STO and $PbTiO_3$ (PTO) [superlattices of $(STO)_n/(PTO)_n$ grown on 5nm $SrRuO_3$ (SRO) buffered $DyScO_3$ (DSO)]. In all cases, prominent superstructure scattering rods corresponding to real-space d-spacings of 9-11 nm have been observed, which are thought to arise from vortex-antivortex pairs of ferroelectric PTO. Without the need to understand the complex physics behind the vortex structures and considering the data provided by V. A. Stoica, et.al. [40] (showing that a single laser pump induces metastable ferroelectric domains with long-range order spanning over all three crystalline dimensions), we can now correlate the observations of these scattering rods within the framework of a simple Moiré interference. In this sample system, PTO grows isomorph (yet slightly strained) on pseudocubic SRO ($a_{SRO}$ = 3.923 Å) in one direction, yielding PTO in-plane lattice constants of $a_{PTO}$ = 3.929 Å and $c_{PTO}$ = 4.099 Å (note, that in-plane PTO polarization is favored by electrostatic alignment with slightly strained STO [40]). Interference of the $c_{PTO}$ dimension with the strong-scattering DSO substrate (pseudocubic $a_{DSO}$= 3.944 Å) then yields a Moiré constant of 10.4 nm, which is in outstanding agreement with previous observations of scattering rods in the $q_x$-$q_z$-plane [37–40]. In the normal direction, interference of the $a_{PTO}$ dimensions with fully relaxed STO ($a_{STO}$ = 3.905 Å) yields a Moiré constant of 64.0 nm, which corresponds well with the easily overlooked scattering-rods in the $q_x$-$q_z$-plane corresponding to a d-spacing of 63.8 nm (found on the STO and PTO (0 0 4) reflections of the high-quality dataset in Fig. 1a of V. A. Stoica, et.al. [40]). Considering the mismatch between the contributing lattices hence offers a simple route to predict resulting vortex superlattice dimensions, needless of complex approaches for equilibration of electrostatic interactions. The fact that STO shows Moiré lattices for more than one substrate indicates that it is the STO surface that acts a lubricant allowing adjacent layers to relax. Even though the interface is crucial, the interference between the two lattices required to obtain this Moiré-motif does not need to be originated by a simple corrugation as for our system, it can create interesting orbital vortex structures.

In summary, we observe near-perfect relaxation of a thin STO film grown on a low lattice-mismatch LSAT substrate, leading to a 2D-square Moiré-pattern. The resulting Moiré domains with approximately 40 nm relate to a commensurate 106/107 supercell relation – an unprecedented and extraordinary observation, as the here found periodicity is 3-4 times larger compared to reported (few-layer) systems. This perfect lattice relaxation implies a long-range (but not short-range) ordered bonding structure at the hetero-interface, as oxygen-sharing between LSAT and STO is only feasible when the corresponding unit cells overlap. Such a locally disordered interface constitutes the ideal environment for gas- and glass-like phases such as 2D electron gas systems: the possibility to detect and characterize the atomic order at the interface, particularly over macroscopic length-scales, will be of significant interest for design and optimization of such promising hetero-structured material systems. Regarding the recent and exciting observation of 3D ferroelectric vortex structures and polar supercrystals, our explanation now allows prediction of expected polar domain sizes simply based on the atomic lattice, which will greatly facilitate the design of new two and three dimensional ferroelectric vortex geometries. Furthermore, Moiré superlattices are known to induce collective phenomena (such as e.g. new plasmon modes) in the electronic landscape, which might open a range of unforeseen potential applications of the LSAT/STO material system.

**AUTHOR INFORMATION**

**Corresponding Authors**

* Urs Staub, email: urs.staub@psi.ch

* Max Burian, email: max.burian@psi.ch

**ACKNOWLEDGMENT**

We acknowledge P. Willmott for helpful discussions and C. Schlepütz for input in data treatment and visualization as well as the use of the PSI SPM Userlab. We further thank V.A.Stoica, V.Gopalan and J.W.Freeland for insightful discussions. M.B. and N.O.H. were supported by the Swiss National Science Foundations Project No. 200021-169017 and H. U. by the National Center of Competence in Research in Molecular Ultrafast Science and Technology (NCCR MUST) from the Swiss National Sciences Foundations. M. R. acknowledges the support by the Swiss National Science Foundations, Project No. 200021−182695.

**REFERENCES**

[1]   J. Mannhart and D. G. Schlom, Science **327**, 1607 (2010).

[2]   M. J. Jin, S. Y. Moon, J. Park, V. Modepalli, J. Jo, S. I. Kim, H. C. Koo, B. C. Min, H. W. Lee, S. H. Baek, and J. W. Yoo, Nano Lett. **17**, 36 (2017).

[3]   M. Gibert, M. Viret, A. Torres-Pardo, C. Piamonteze, P. Zubko, N. Jaouen, J. M. Tonnerre, A. Mougin, J. Fowlie, S. Catalano, A. Gloter, O. Stéphan, and J. M. Triscone, Nano Lett. **15**, 7355 (2015).

[4]   S. R. Lee, L. Baasandorj, J. W. Chang, I. W. Hwang, J. R. Kim, J. G. Kim, K. T. Ko, S. B. Shim, M. W. Choi, M. You, C. H. Yang, J. Kim, and J. Song, Nano Lett. **19**, 2243 (2019).

[5] T. L. Kim, M. J. Choi, T. H. Lee, W. Sohn, and H. W. Jang, Nano Lett. **19**, 5897 (2019).

[6] C. W. Bark, P. Sharma, Y. Wang, S. H. Baek, S. Lee, S. Ryu, C. M. Folkman, T. R. Paudel, A. Kumar, S. V. Kalinin, A. Sokolov, E. Y. Tsymbal, M. S. Rzchowski, A. Gruverman, and C. B. Eom, Nano Lett. **12**, 1765 (2012).

[7] L. Wu, C. Li, M. Chen, Y. Zhang, K. Han, S. Zeng, X. Liu, J. Ma, C. Liu, J. Chen, J. Zhang, Ariando, T. V. Venkatesan, S. J. Pennycook, J. M. D. Coey, L. Shen, J. Ma, X. R. Wang, and C. W. Nan, ACS Appl. Mater. Interfaces **9**, 44931 (2017).

[8] C. A. F Vaz, F. J. Walker, C. H. Ahn, and S. Ismail-Beigi, J. Phys. Condens. Matter 123001 (2015).

[9] H. J. A. Molegraaf, J. Hoffman, C. A. F. Vaz, S. Gariglio, D. van der Marel, C. H. Ahn, and J.-M. Triscone, Adv. Mater. **21**, 3470 (2009).

[10] S.-W. Cheong and M. Mostovoy, Nat. Mater. **6**, 13 (2007).

[11] C. Rao, Annu. Rev. Phys. Chem. **40**, 291 (1989).

[12] N. Reyren, S. Thiel, A. D. Caviglia, L. F. Kourkoutis, G. Hammerl, C. Richter, C. W. Schneider, T. Kopp, A.-S. Ruetschi, D. Jaccard, M. Gabay, D. A. Muller, J.-M. Triscone, and J. Mannhart, Science **317**, 1196 (2007).

[13] M. Huijben, G. Rijnders, D. H. A. Blank, S. Bals, S. Van Aert, J. Verbeeck, G. Van Tendeloo, A. Brinkman, and H. Hilgenkamp, Nat. Mater. **5**, 556 (2006).

[14] A. Ohtomo and H. Y. Hwang, Nature **427**, 423 (2004).

[15] Y. Chen, R. J. Green, R. Sutarto, F. He, S. Linderoth, G. A. Sawatzky, and N. Pryds, Nano Lett. **17**, 7062 (2017).

[16] M. A. Islam, D. Saldana-Greco, Z. Gu, F. Wang, E. Breckenfeld, Q. Lei, R. Xu, C. J. Hawley, X.

X. Xi, L. W. Martin, A. M. Rappe, and J. E. Spanier, Nano Lett. **16**, 681 (2016).

[17]  W. Niu, Y. Zhang, Y. Gan, D. V. Christensen, M. V. Soosten, E. J. Garcia-Suarez, A. Riisager, X. Wang, Y. Xu, R. Zhang, N. Pryds, and Y. Chen, Nano Lett. **17**, 6878 (2017).

[18]  Y. Xie, C. Bell, T. Yajima, Y. Hikita, and H. Y. Hwang, Nano Lett. **10**, 2588 (2010).

[19]  Y. Z. Chen, J. L. Zhao, J. R. Sun, N. Pryds, and B. G. Shen, Appl. Phys. Lett. **97**, 123102 (2010).

[20]  P. Zubko, N. Jecklin, A. Torres-Pardo, P. Aguado-Puente, A. Gloter, C. Lichtensteiger, J. Junquera, O. Stéphan, and J.-M. Triscone, Nano Lett. **12**, 2846 (2012).

[21]  A. Brinkman, M. Huijben, M. van Zalk, J. Huijben, U. Zeitler, J. C. Maan, W. G. van der Wiel, G. Rijnders, D. H. A. Blank, and H. Hilgenkamp, Nat. Mater. **6**, 493 (2007).

[22]  H. Ohta, S. Kim, Y. Mune, T. Mizoguchi, K. Nomura, S. Ohta, T. Nomura, Y. Nakanishi, Y. Ikuhara, M. Hirano, H. Hosono, and K. Koumoto, Nat. Mater. **6**, 129 (2007).

[23]  J. Garcia-Barriocanal, A. Rivera-Calzada, M. Varela, Z. Sefrioui, E. Iborra, C. Leon, S. J. Pennycook, and J. Santamaria, Science **321**, 676 (2008).

[24]  A. Soumyanarayanan, N. Reyren, A. Fert, and C. Panagopoulos, Nature **539**, 509 (2016).

[25]  A. Manchon, H. C. Koo, J. Nitta, S. M. Frolov, and R. A. Duine, Nat. Mater. **14**, 871 (2015).

[26]  V. Sunko, H. Rosner, P. Kushwaha, S. Khim, F. Mazzola, L. Bawden, O. J. Clark, J. M. Riley, D. Kasinathan, M. W. Haverkort, T. K. Kim, M. Hoesch, J. Fujii, I. Vobornik, A. P. Mackenzie, and P. D. C. King, Nature **549**, 492 (2017).

[27]  W. Lin, L. Li, F. Doğan, C. Li, H. Rotella, X. Yu, B. Zhang, Y. Li, W. S. Lew, S. Wang, W. Prellier, S. J. Pennycook, J. Chen, Z. Zhong, A. Manchon, and T. Wu, Nat. Commun. **10**, (2019).

[28]  C. R. Dean, L. Wang, P. Maher, C. Forsythe, F. Ghahari, Y. Gao, J. Katoch, M. Ishigami, P. Moon,

M. Koshino, T. Taniguchi, K. Watanabe, K. L. Shepard, J. Hone, and P. Kim, Nature **497**, 598 (2013).

[29] K. Tran, G. Moody, F. Wu, X. Lu, J. Choi, K. Kim, A. Rai, D. A. Sanchez, J. Quan, A. Singh, J. Embley, A. Zepeda, M. Campbell, T. Autry, T. Taniguchi, K. Watanabe, N. Lu, S. K. Banerjee, K. L. Silverman, S. Kim, E. Tutuc, L. Yang, A. H. MacDonald, and X. Li, Nature **567**, 71 (2019).

[30] J. L. Servoin, Y. Luspin, and F. Gervais, Phys. Rev. B **22**, 5501 (1980).

[31] A. A. Sirenko, C. Bernhard, A. Golnik, A. M. Clark, J. Hao, W. Si, and X. X. Xi, Nature **404**, 373 (2000).

[32] P. Marsik, K. Sen, J. Khmaladze, M. Yazdi-Rizi, B. P. P. Mallett, and C. Bernhard, Appl. Phys. Lett. **108**, 052901 (2016).

[33] Y. Yamada and G. Shirane, J. Phys. Soc. Japan **26**, 396 (1969).

[34] J. H. Haeni, P. Irvin, W. Chang, R. Uecker, P. Reiche, Y. L. Li, S. Choudhury, W. Tian, M. E. Hawley, B. Craigo, A. K. Tagantsev, X. Q. Pan, S. K. Streiffer, L. Q. Chen, S. W. Kirchoefer, J. Levy, and D. G. Schlom, Nature **430**, 758 (2004).

[35] V. Skoromets, C. Kadlec, J. Drahokoupil, J. Schubert, J. Hlinka, and P. Kužel, Phys. Rev. B - Condens. Matter Mater. Phys. **89**, (2014).

[36] A. Verma, S. Raghavan, S. Stemmer, and D. Jena, Appl. Phys. Lett. **107**, (2015).

[37] A. R. Damodaran, J. D. Clarkson, Z. Hong, H. Liu, A. K. Yadav, C. T. Nelson, S. L. Hsu, M. R. McCarter, K. D. Park, V. Kravtsov, A. Farhan, Y. Dong, Z. Cai, H. Zhou, P. Aguado-Puente, P. Garcia-Fernandez, J. Iniguez, J. Junquera, A. Scholl, M. B. Raschke, L. Q. Chen, D. D. Fong, R. Ramesh, and L. W. Martin, Nat. Mater. **16**, 1003 (2017).

[38] A. K. Yadav, C. T. Nelson, S. L. Hsu, Z. Hong, J. D. Clarkson, C. M. Schlepuëtz, A. R. Damodaran,

P. Shafer, E. Arenholz, L. R. Dedon, D. Chen, A. Vishwanath, A. M. Minor, L. Q. Chen, J. F. Scott, L. W. Martin, and R. Ramesh, Nature **530**, 198 (2016).

[39] S. Das, Y. L. Tang, Z. Hong, M. A. P. Gonçalves, M. R. McCarter, C. Klewe, K. X. Nguyen, F. Gómez-Ortiz, P. Shafer, E. Arenholz, V. A. Stoica, S. L. Hsu, B. Wang, C. Ophus, J. F. Liu, C. T. Nelson, S. Saremi, B. Prasad, A. B. Mei, D. G. Schlom, J. Íñiguez, P. García-Fernández, D. A. Muller, L. Q. Chen, J. Junquera, L. W. Martin, and R. Ramesh, Nature **568**, 368 (2019).

[40] V. A. Stoica, N. Laanait, C. Dai, Z. Hong, Y. Yuan, Z. Zhang, S. Lei, M. R. McCarter, A. Yadav, A. R. Damodaran, S. Das, G. A. Stone, J. Karapetrova, D. A. Walko, X. Zhang, L. W. Martin, R. Ramesh, L.-Q. Chen, H. Wen, V. Gopalan, and J. W. Freeland, Nat. Mater. **18**, 377 (2019).

[41] S. W. Lee, Y. Liu, J. Heo, and R. G. Gordon, Nano Lett. **12**, 4775 (2012).

[42] M. Steins, J. Doerschel, and P. Reiche, Zeitschrift Fur Krist. - New Cryst. Struct. **212**, 77 (1997).

[43] M. Kozina, M. Fechner, P. Marsik, T. van Driel, J. M. Glownia, C. Bernhard, M. Radovic, D. Zhu, S. Bonetti, U. Staub, and M. C. Hoffmann, Nat. Phys. **15**, 387 (2019).

[44] Z. Huang, K. Han, S. Zeng, M. Motapothula, A. Y. Borisevich, S. Ghosh, W. Lü, C. Li, W. Zhou, Z. Liu, M. Coey, T. Venkatesan, and Ariando, Nano Lett. **16**, 2307 (2016).

[45] P. R. Willmott, D. Meister, S. J. Leake, M. Lange, A. Bergamaschi, M. Böge, M. Calvi, C. Cancellieri, N. Casati, A. Cervellino, Q. Chen, C. David, U. Flechsig, F. Gozzo, B. Henrich, S. Jäggi-Spielmann, B. Jakob, I. Kalichava, P. Karvinen, J. Krempasky, A. Lüdeke, R. Lüscher, S. Maag, C. Quitmann, M. L. Reinle-Schmitt, T. Schmidt, B. Schmitt, A. Streun, I. Vartiainen, M. Vitins, X. Wang, and R. Wullschleger, J. Synchrotron Radiat. **20**, 667 (2013).

[46] B. D. Patterson, R. Abela, H. Auderset, Q. Chen, F. Fauth, F. Gozzo, G. Ingold, H. Kühne, M. Lange, D. Maden, D. Meister, P. Pattison, T. Schmidt, B. Schmitt, C. Schulze-Briese, M. Shi, M.

Stampanoni, and P. R. Willmott, Nucl. Instruments Methods Phys. Res. Sect. A Accel. Spectrometers, Detect. Assoc. Equip. **540**, 42 (2005).

[47] C. M. Schlepütz, S. O. Mariager, S. A. Pauli, R. Feidenhans'l, and P. R. Willmott, J. Appl. Crystallogr. **44**, 73 (2011).

[48] E.-J. Guo, R. Desautels, D. Keavney, M. A. Roldan, B. J. Kirby, D. Lee, Z. Liao, T. Charlton, A. Herklotz, T. Zac Ward, M. R. Fitzsimmons, and H. N. Lee, Sci. Adv. **5**, eaav5050 (2019).

[49] J. G. Connell, B. J. Isaac, G. B. Ekanayake, D. R. Strachan, and S. S. A. Seo, Appl. Phys. Lett. **101**, (2012).

[50] D. Xu, Y. Yuan, H. Zhu, L. Cheng, C. Liu, J. Su, X. Zhang, H. Zhang, X. Zhang, and J. Li, Materials (Basel). **12**, (2019).

[51] M. Radovic, N. Lampis, F. M. Granozio, P. Perna, Z. Ristic, M. Salluzzo, C. M. Schlepütz, and U. S. Di Uccio, Appl. Phys. Lett. **94**, 022901 (2009).

[52] P. P. Ewald, Acta Crystallogr. Sect. A Cryst. Physics, Diffraction, Theor. Gen. Crystallogr. **25**, 103 (1969).

[53] C. Nicklin, Science **343**, 739 (2014).

[54] A. Neagu and C.-W. Tai, Sci. Rep. **7**, 12519 (2017).

[55] G. X. Ni, H. Wang, J. S. Wu, Z. Fei, M. D. Goldflam, F. Keilmann, B. Özyilmaz, A. H. Castro Neto, X. M. Xie, M. M. Fogler, and D. N. Basov, Nat. Mater. **14**, 1217 (2015).

[56] D. Martoccia, T. Brugger, M. Björck, C. M. Schlepütz, S. A. Pauli, T. Greber, B. D. Patterson, and P. R. Willmott, Surf. Sci. **604**, (2010).

[57] P. Zeller, X. Ma, and S. Günther, New J. Phys. **19**, (2017).

[58] P. Merino, M. Švec, A. L. Pinardi, G. Otero, and J. A. Martín-Gago, in *ACS Nano* (2011), pp. 5627–5634.

[59] J. Narayan and B. C. Larson, J. Appl. Phys. **93**, 278 (2003).

[60] M. Iannuzzi, I. Kalichava, H. Ma, S. J. Leake, H. Zhou, G. Li, Y. Zhang, O. Bunk, H. Gao, J. Hutter, P. R. Willmott, and T. Greber, Phys. Rev. B **88**, 125433 (2013).

[61] D. Martoccia, M. Björck, C. M. Schlepütz, T. Brugger, S. A. Pauli, B. D. Patterson, T. Greber, and P. R. Willmott, New J. Phys. **12**, 43028 (2010).